COLLEGE OF ELECTRICAL AND MECHANICAL ENGINEERING

NATIONAL UNIVERSITY OF SCIENCES AND TECHNOLOGY (NUST)

# RISC AND CISC

## Computer Architecture

**By**

**Farhat Masood**
**BE – Electrical (NUST)**

# TABLE OF CONTENTS

## LIST OF FIGURES

# RISC AND CISC

## General[i]

1. The dominant architecture in the PC market, the Intel IA-32, belongs to the Complex Instruction Set Computer (CISC) design. The obvious reason for this classification is the “complex” nature of its Instruction Set Architecture (ISA). The motivation for designing such complex instruction sets is to provide an instruction set that closely supports the operations and data structures used by Higher-Level Languages (HLLs). However, the side effects of this design effort are far too serious to ignore.

## Addressing Modes in CISC

2. The decision of CISC processor designers to provide a variety of addressing modes leads to variable-length instructions. For example, instruction length increases if an operand is in memory as opposed to in a register.

   a. This is because we have to specify the memory address as part of instruction encoding, which takes many more bits.

   b. This complicates instruction decoding and scheduling. The side effect of providing a wide range of instruction types is that the number of clocks required to execute instructions varies widely.

   c. This again leads to problems in instruction scheduling and pipelining.

## Evolution of RISC[ii]

3. For these and other reasons, in the early 1980s designers started looking at simple ISAs. Because these ISAs tend to produce instruction sets with far fewer instructions, they coined the term Reduced Instruction Set Computer (RISC). Even though the main goal was not to reduce the number of instructions, but the complexity, the term has stuck.

4. There is no precise definition of what constitutes a RISC design. However, we can identify certain characteristics that are present in most RISC systems.

a. We identify these RISC design principles after looking at why the designers took the route of CISC in the first place.

b. Because CISC and RISC have their advantages and disadvantages, modern processors take features from both classes. For example, the PowerPC, which follows the RISC philosophy, has quite a few complex instructions.

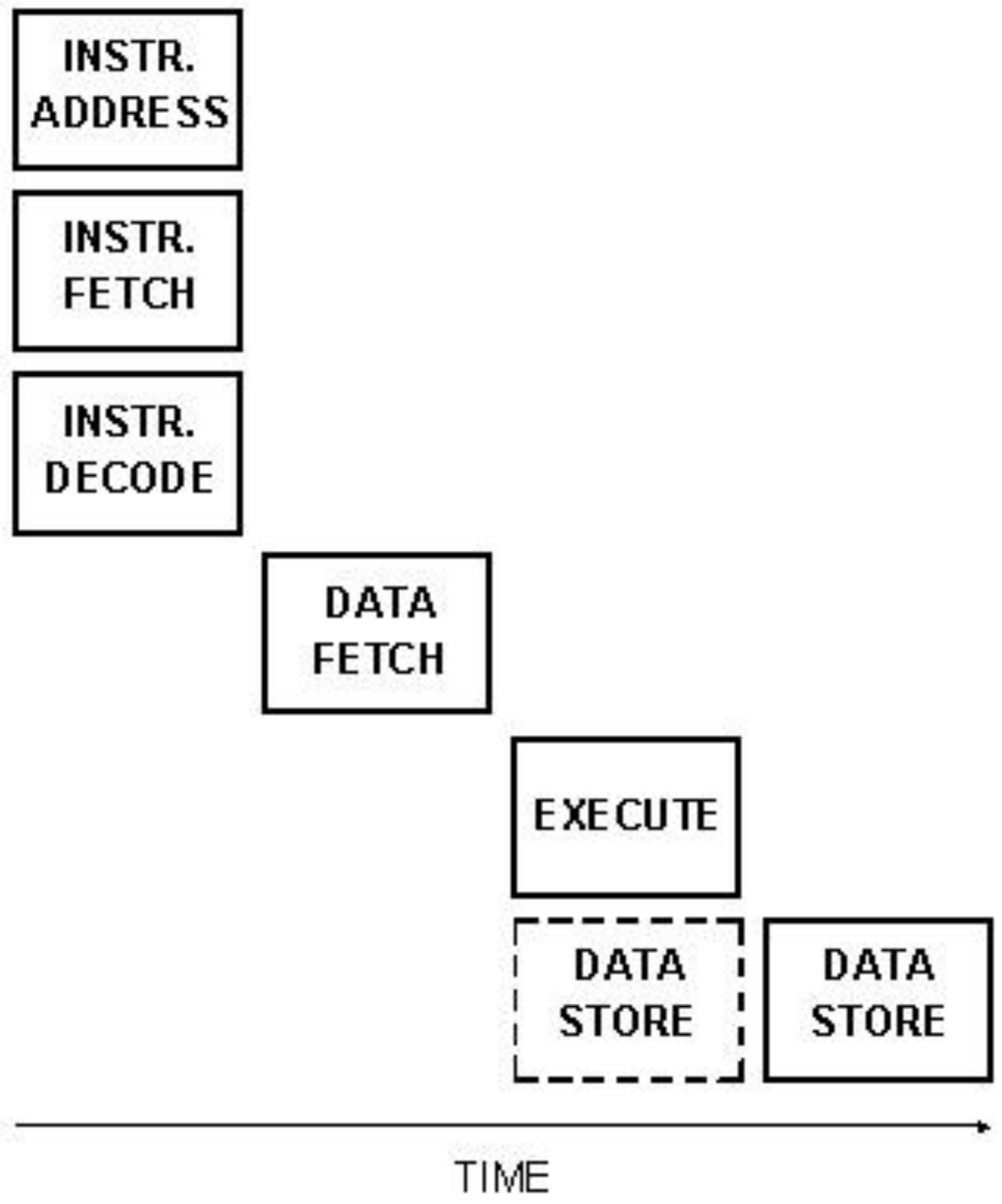

Figure 1 **<u>Typical RISC Architecture based Machine - Instruction phase overlapping</u>**

**<u>Definition of RISC[iii]</u>**

5. RISC, or *Reduced Instruction Set Computer* is a type of microprocessor architecture that utilizes a small, highly-optimized set of instructions, rather than a more specialized set of instructions often found in other types of architectures.

a. **<u>Evolution/History</u>**. The first RISC projects came from IBM, Stanford, and UC-Berkeley in the late 70s and early 80s. The IBM 801, Stanford

MIPS, and Berkeley RISC 1 and 2 were all designed with a similar philosophy which has become known as RISC. Certain design features have been characteristic of most RISC processors

(1) **One Cycle Execution Time**. RISC processors have a CPI (clock per instruction) of one cycle. This is due to the optimization of each instruction on the CPU and a technique called ;

(2) **Pipelining**. A technique that allows for simultaneous execution of parts, or stages, of instructions to more efficiently process instructions;

(3) **Large Number of Registers**. The RISC design philosophy generally incorporates a larger number of registers to prevent in large amounts of interactions with memory

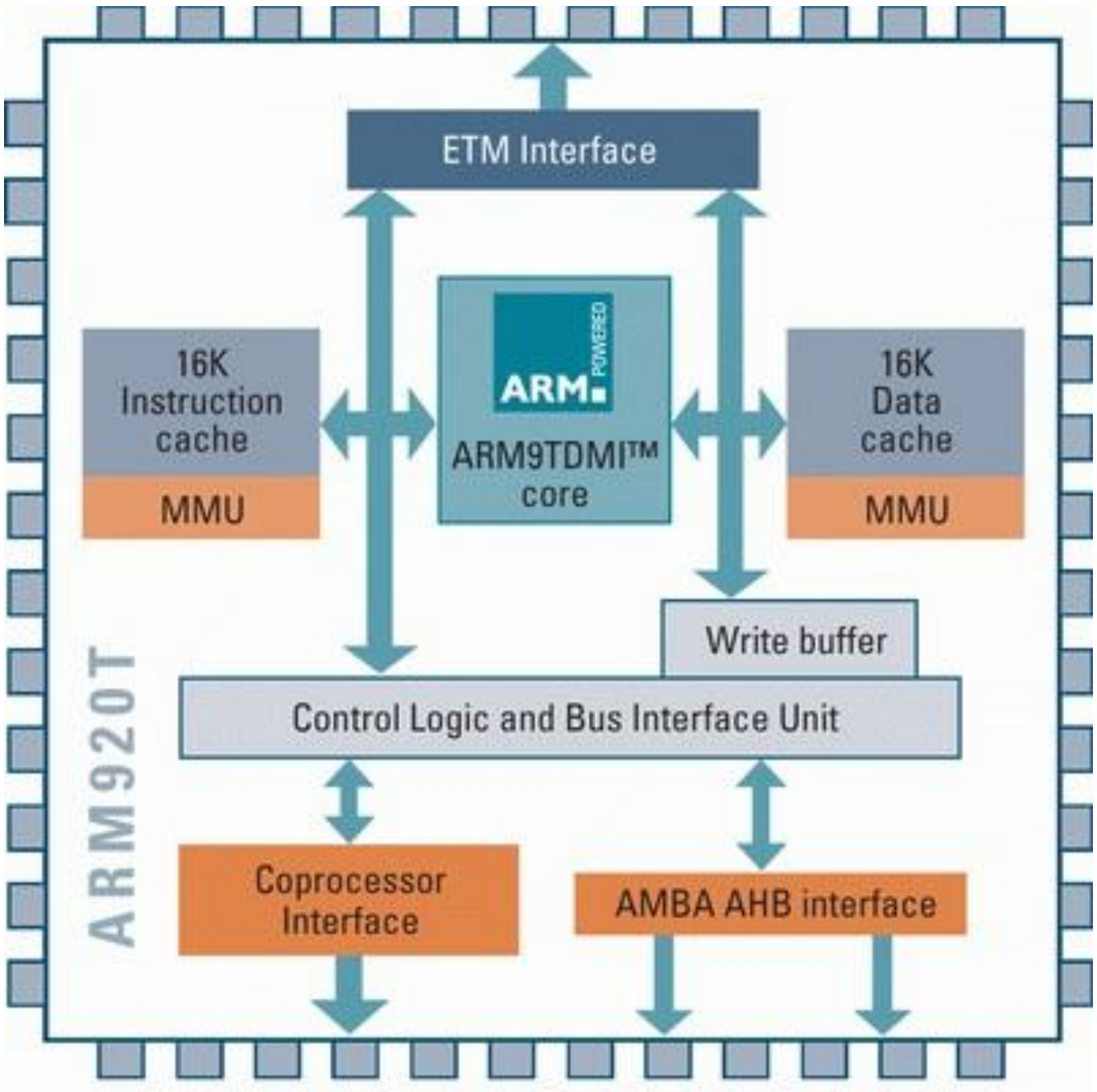

Figure 2 **Advanced RISC Machine (ARM)**

**Non RISC Design or Pre RISC Design**[iv]

6. In the early days of the computer industry, programming was done in assembly language or machine code, which encouraged powerful and easy to use instructions. CPU designers therefore tried to make instructions that would do as much work as possible. With the advent of higher level languages, computer architects also started to create dedicated instructions to directly implement certain central mechanisms of such languages.

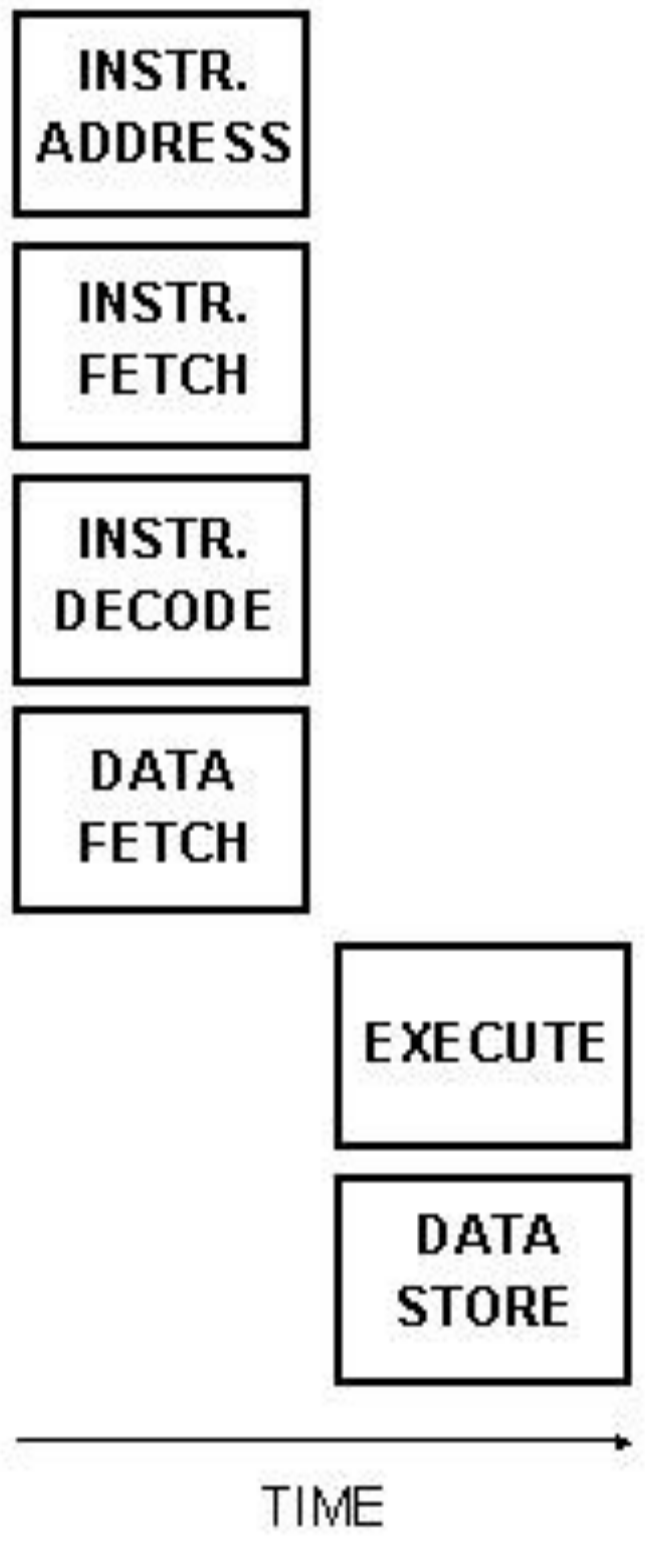

Figure 3 **Typical CISC Architecture – Stack Design**

7. Another general goal was to provide every possible addressing mode for every instruction, known as orthogonality, to ease compiler implementation. Arithmetic operations could therefore often have results as well as operands directly in memory (in addition to register or immediate).

8. The attitude at the time was that hardware design was more mature than compiler design so this was in itself also a reason to implement parts of the functionality in hardware and/or microcode rather than in a memory constrained compiler (or its generated code) alone. This design philosophy became retroactively termed Complex Instruction Set Computer (CISC) after the RISC philosophy came onto the scene.

9. An important force encouraging complexity was very limited main memories (on the order of kilobytes). It was therefore advantageous for the density of information held in computer programs to be high, leading to features such as highly encoded, variable length instructions, doing data loading as well as. These issues were of higher priority than the ease of decoding such instructions.

10. An equally important reason was that main memories were quite slow (a common type was ferrite core memory); by using dense information packing, one could reduce the frequency with which the CPU had to access this slow resource. Modern computers face similar limiting factors: main memories are slow compared to the CPU and the fast cache memories employed to overcome this are instead limited in size. This may partly explain why highly encoded instruction sets have proven to be as useful as RISC designs in modern computers.

**Typical Characteristics of RISC Architecture**[v]

11. Designers make choices based on the available technology. As the technology, both hardware and software, evolves, design choices also evolve. Furthermore, as we get more experience in designing processors, we can design better systems. The RISC proposal was a response to the changing technology and the accumulation of knowledge from the CISC designs. CISC processors were designed to simplify compilers and to improve performance under constraints such as small and slow memories. The important observations that motivated designers to consider alternatives to CISC designs were

a. **Simple Instructions**. The designers of CISC architectures anticipated extensive use of complex instructions because they close the semantic gap. In reality, it turns out that compilers mostly ignore these instructions. Several empirical studies have shown that this is the case. One reason for this is that different high-level languages use different semantics. For example, the semantics of the C for loop is not exactly the same as that in other languages. Thus, compilers tend to synthesize the code using simpler instructions.

b. **Few Data Types**. CISC ISA tends to support a variety of data structures, from simple data types such as integers and characters to complex data structures such as records and structures. Empirical data suggest that complex data structures are used relatively infrequently. Thus, it is beneficial to design a system that supports a few simple data types efficiently and from which the missing complex data types can be synthesized.

c. **Simple Addressing Modes**. CISC designs provide a large number of addressing modes. The main motivations are

(1) To support complex data structures and

(2) To provide flexibility to access operands.

(a) Problems Caused. Although this allows flexibility, it also introduces problems. First, it causes variable instruction execution times, depending on the location of the operands.

(b) Second, it leads to variable-length instructions. For example, the IA-32 instruction length can range from 1 to 12 bytes. Variable instruction lengths lead to inefficient instruction decoding and scheduling.

d. **Identical General Purpose Registers**. Allowing any register to be used in any context, simplifying compiler design (although normally there are separate floating point registers).

e. **Harvard Architecture Based**. RISC designs are also more likely to feature a Harvard memory model, where the instruction stream and the data stream are conceptually separated; this means that modifying the memory where code is held might not have any effect on the instructions executed by the processor (because the CPU has a separate instruction and data cache), at least until a special synchronization instruction is issued. On the upside, this allows both caches to be accessed simultaneously, which can often improve performance.

**RISC VS CISC – An Example**[vi]

12. The simplest way to examine the advantages and disadvantages of RISC architecture is by contrasting it with its predecessor, CISC (Complex Instruction Set Computers) architecture.

13. **Multiplying Two Numbers in Memory**. The main memory is divided into locations numbered from (row) 1: (column) 1 to (row) 6: (column) 4. The execution unit is responsible for carrying out all computations. However, the execution unit can only operate on data that has been loaded into one of the six registers (A, B, C, D, E, or F). Let's say we want to find the product of two numbers - one stored in location 2:3 and another stored in location 5:2 - and then store the product back in the location 2:3

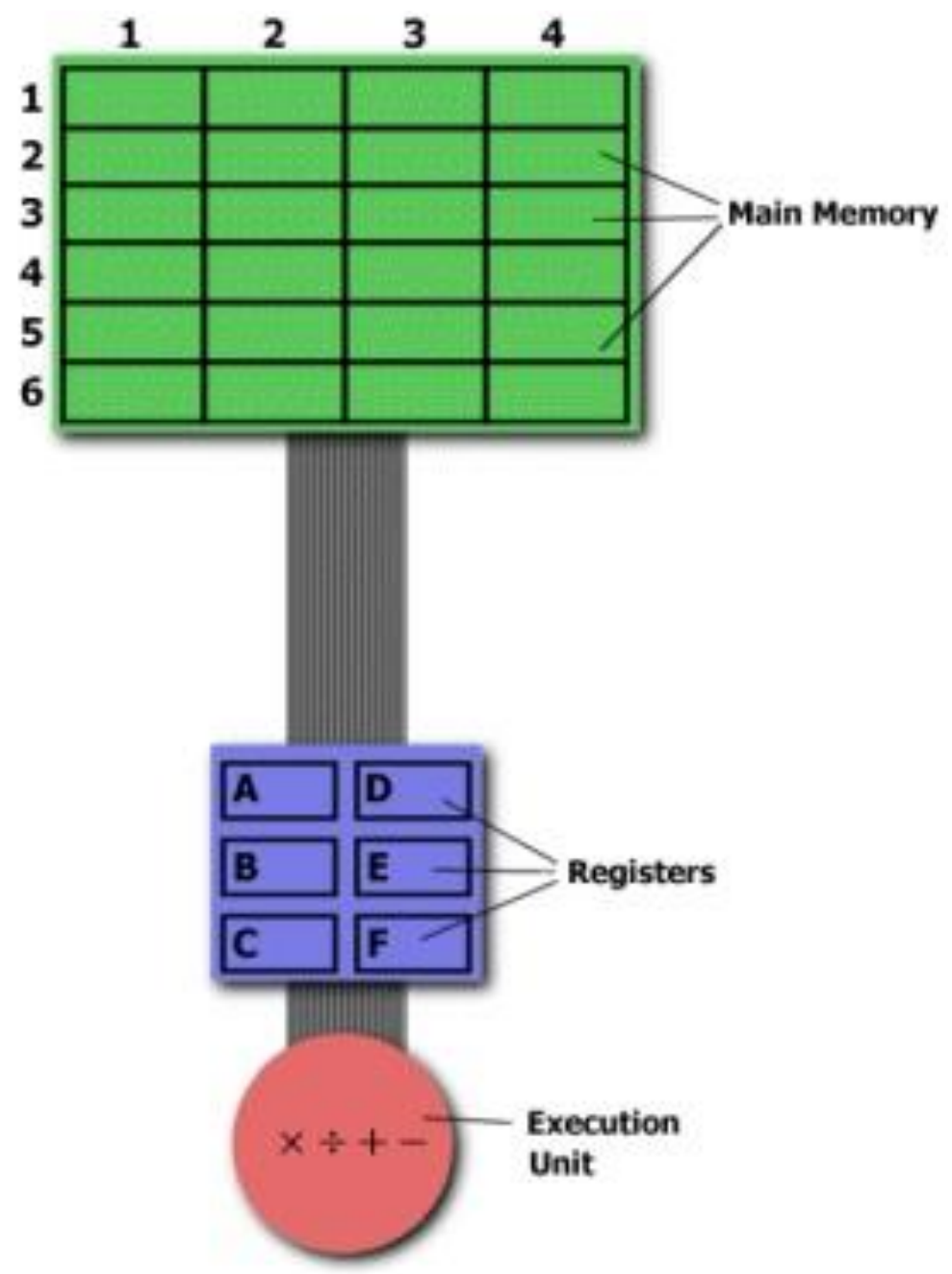

Figure 4 **Representation of Storage Scheme for a Generic Computer**

14. **The CISC Approach**. The primary goal of CISC architecture is to complete a task in as few lines of assembly as possible. This is achieved by building processor hardware that is capable of understanding and executing a series of operations. For this particular task, a CISC processor would come prepared with a specific instruction (say "MUL").

a. When executed, this instruction loads the two values into separate registers, multiplies the operands in the execution unit, and then stores the product in the appropriate register.

b. Thus, the entire task of multiplying two numbers can be completed with one instruction:

*MUL 2:3, 5:2*

c. MUL is what is known as a "complex instruction."

d. It operates directly on the computer's memory banks and does not require the programmer to explicitly call any loading or storing functions.

e. It closely resembles a command in a higher level language. For instance, if we let "a" represent the value of 2:3 and "b" represent the value of 5:2, then this command is identical to the C statement "a = a x b."

15. Advantage. One of the primary advantages of this system is that the compiler has to do very little work to translate a high-level language statement into assembly. Because the length of the code is relatively short, very little RAM is required to store instructions. The emphasis is put on building complex instructions directly into the hardware.

16. **The RISC Approach**. RISC processors only use simple instructions that can be executed within one clock cycle. Thus, the "MUL" command described above could be divided into three separate commands:

a. "LOAD," which moves data from the memory bank to a register,

b. "PROD," which finds the product of two operands located within the registers, and

c. "STORE," which moves data from a register to the memory banks.

d. In order to perform the exact series of steps described in the CISC approach, a programmer would need to code four lines of assembly:

*LOAD A, 2:3*
*LOAD B, 5:2*
*PROD A, B*
*STORE 2:3, A*

17. **Analysis**. At first, this may seem like a much less efficient way of completing the operation. Because there are more lines of code, more RAM is needed to store the assembly level instructions. The compiler must also perform more work to convert a high-level language statement into code of this form.

a. Advantage of RISC. However, the RISC strategy also brings some very important advantages. Because each instruction requires only one clock cycle to execute, the entire program will execute in approximately the same amount of time as the multi-cycle "MUL" command. These RISC "reduced instructions" require less transistors of hardware space than the complex instructions, leaving more room for general purpose registers. Because all of the instructions execute in a uniform amount of time (i.e. one clock), pipelining is possible.

(1) Separating the "LOAD" and "STORE" instructions actually reduces the amount of work that the computer must perform.

(2) After a CISC-style "MUL" command is executed, the processor automatically erases the registers. If one of the operands needs to be used for another computation, the processor must re-load the data from the memory bank into a register. In RISC, the operand will remain in the register until another value is loaded in its place.

b. The following table will differentiate both the architectures and based on the analysis the overall advantage will be discussed.

| CISC | RISC |
|---|---|
| Emphasis on hardware | Emphasis on software |
| Includes multi-clock complex instructions | Single-clock, reduced instruction only |
| Memory-to-memory: "LOAD" and "STORE" incorporated in instructions | Register to register: "LOAD" and "STORE" are independent instructions |
| Small code sizes, high cycles per second | Low cycles per second, large code sizes |
| Transistors used for storing complex instructions | Spends more transistors on memory registers |

Table 1 **Comparison of CISC and RISC Architectures**[vii]

18. **The Performance Equation**. The following equation is commonly used for expressing a computer's performance ability:

$$\frac{\text{time}}{\text{program}} = \frac{\text{time}}{\text{cycle}} \times \frac{\text{cycles}}{\text{instruction}} \times \frac{\text{instructions}}{\text{program}}$$

a. CISC Approach. The CISC approach attempts to minimize the number of instructions per program, sacrificing the number of cycles per instruction.

b. RISC Approach. RISC does the opposite, reducing the cycles per instruction at the cost of the number of instructions per program.

19. **The Overall RISC Advantage**. Today, the Intel x86 is arguable the only chip which retains CISC architecture. This is primarily due to advancements in other areas of computer technology.

a. The price of RAM has decreased dramatically. In 1977, 1MB of DRAM cost about $5,000.

b. By 1994, the same amount of memory cost only $6 (when adjusted for inflation). Compiler technology has also become more sophisticated, so

that the RISC use of RAM and emphasis on software has become ideal.

**RISC Processors (Examples)[viii]**

20. **Digital Equipment Corporation (DEC) - Alpha**.[ix] Alpha, originally known as Alpha AXP, is a 64-bit reduced instruction set computer (RISC) instruction set architecture (ISA) developed by Digital Equipment Corporation (DEC), designed to replace the 32-bit VAX complex instruction set computer (CISC) ISA and its implementations.

a. Alpha was implemented in microprocessors originally developed and fabricated by DEC.

b. These microprocessors were most prominently used in a variety of DEC workstations and servers, which eventually formed the basis for almost their entire mid-to-upper-scale lineup.

c. Several third-party vendors also produced Alpha systems, including PC form factor motherboards.

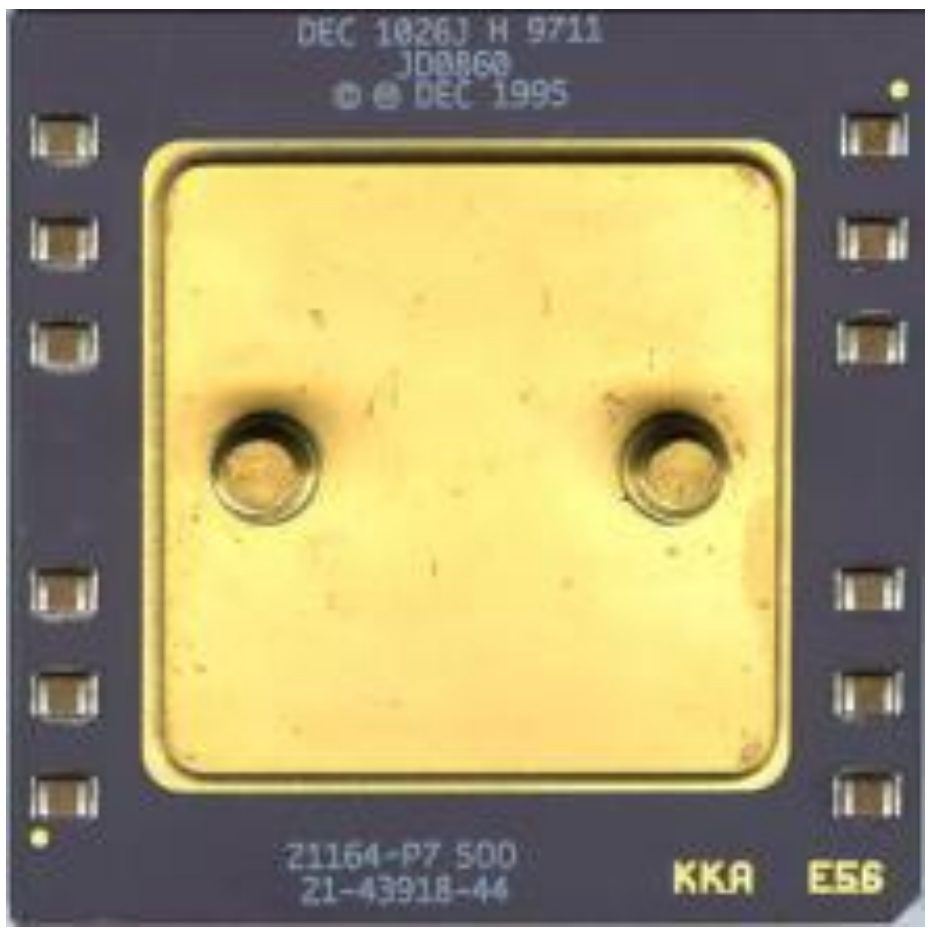

Figure 5 **DEC Alpha Microprocessor developed in 1995**

21. **Advanced Micro Devices (AMD) 29000**.[x] The AMD 29000, often simply 29k, was a popular family of 32-bit RISC microprocessors and microcontrollers developed and fabricated by Advanced Micro Devices (AMD).

a. They were, for a time, the most popular RISC chips on the market, widely used in laser printers from a variety of manufacturers.

b. In late 1995 AMD dropped development of the 29k because the design team was transferred to support the PC side of the business and was realigned towards the embedded 186 family of 80186 derivatives.

c. The majority of AMD's resources were then concentrated on their high-performance, desktop x86 clones, using many of the ideas and individual parts of the latest 29k to produce the AMD K5.

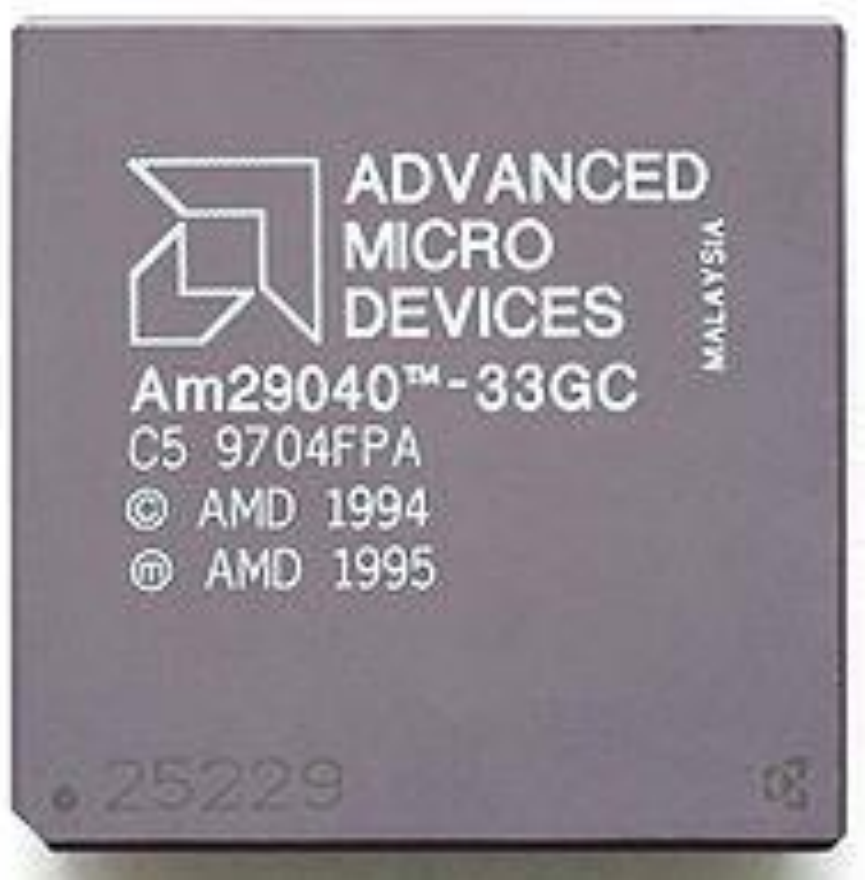

Figure 6 **Advanced Micro Devices (AMD) Microprocessor 29000 Series**

22. **Advanced RISC Machine (ARM)**. The ARM is a 32-bit reduced instruction set computer (RISC) instruction set architecture (ISA) developed by ARM Holdings. It was known as the Advanced RISC Machine, and before that as the Acorn RISC Machine.

a. The ARM architecture is the most widely used 32-bit ISA in terms of numbers produced.

b. They were originally conceived as a processor for desktop personal computers by Acorn Computers, a market now dominated by the x86 family used by IBM PC compatible computers.

c. The relative simplicity of ARM processors made them suitable for low power applications.

d. This has made them dominant in the mobile and embedded electronics market as relatively low cost and small microprocessors and microcontrollers.

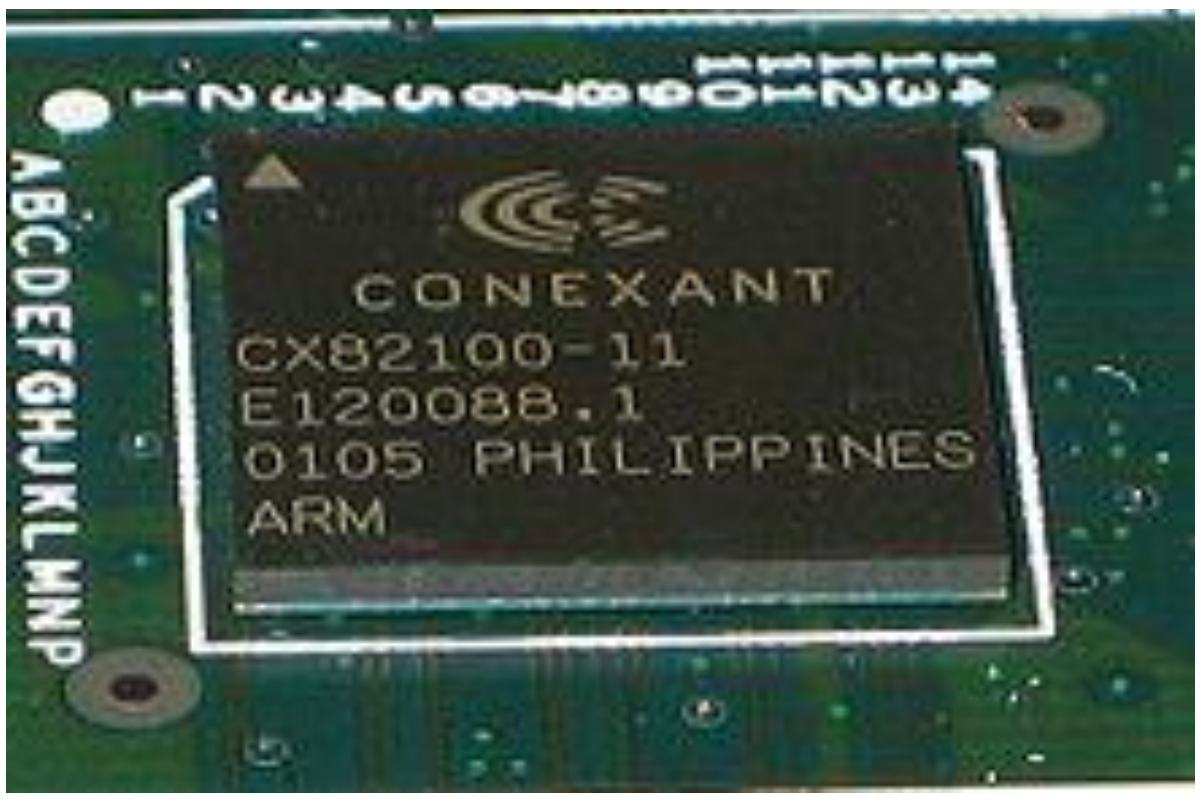

Figure 7 **<u>Advanced RISC Machine (ARM) Microprocessor developed by Conexant Computers</u>**

23. **<u>Atmel AVR[xi]</u>**. The AVR is a Modified Harvard architecture 8-bit RISC single chip microcontroller (µC) which was developed by Atmel in 1996.

a. The AVR was one of the first microcontroller families to use on-chip flash memory for program storage, as opposed to One-Time Programmable ROM, EPROM, or EEPROM used by other microcontrollers at the time.

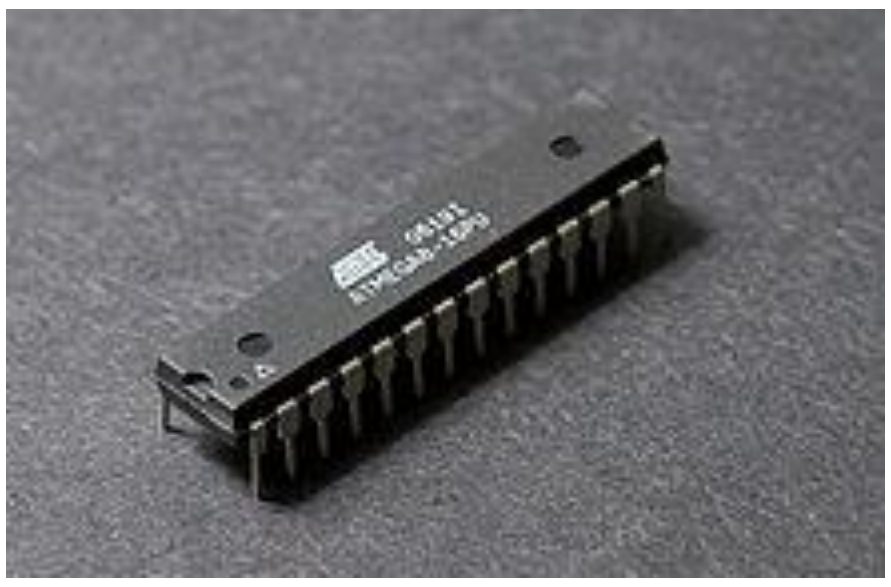

Figure 8 **<u>Atmel AVR ATmega8 Microprocessor</u>**

24. **<u>Microprocessor without Interlocked Pipeline Stages (MIPS[xii])</u>**. MIPS is a reduced instruction set computer (RISC) instruction set architecture (ISA) developed by MIPS Computer Systems.

a. The early MIPS architectures were 32-bit, and later versions were 64-bit.

b. Multiple revisions of the MIPS instruction set exist, including MIPS I, MIPS II, MIPS III, MIPS IV, MIPS V, MIPS32, and MIPS64.

c. The current revisions are MIPS32 (for 32-bit implementations) and MIPS64 (for 64-bit implementations).

d. MIPS32 and MIPS64 define a control register set as well as the instruction set.

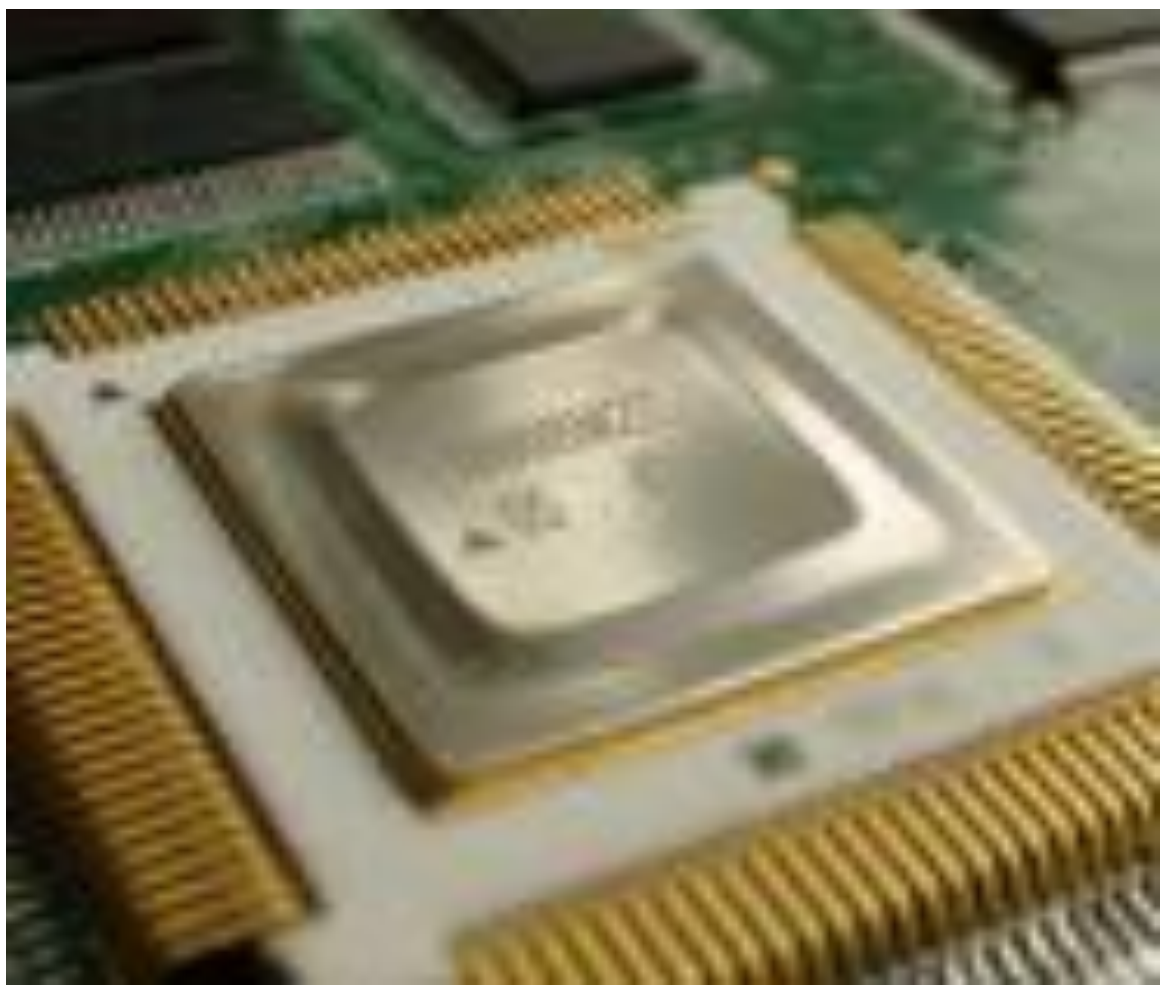

Figure 9 **MIPS Microprocessor**

25. **Precision Architecture – Reduced Instruction Set Computer (PA-RISC)**. [xiii]PA-RISC is an instruction set architecture (ISA) developed by Hewlett-Packard.

a. The design is also referred to as HP/PA for Hewlett Packard Precision Architecture.

b. The architecture comprised of the HP 3000 Series 930 and HP 9000 Model 840 computers.

c. PA-RISC has been succeeded by the Itanium (originally IA-64) ISA jointly developed by HP and Intel.

d. They have stopped the production of HP 9000 series but the support plan is till the year 2013.

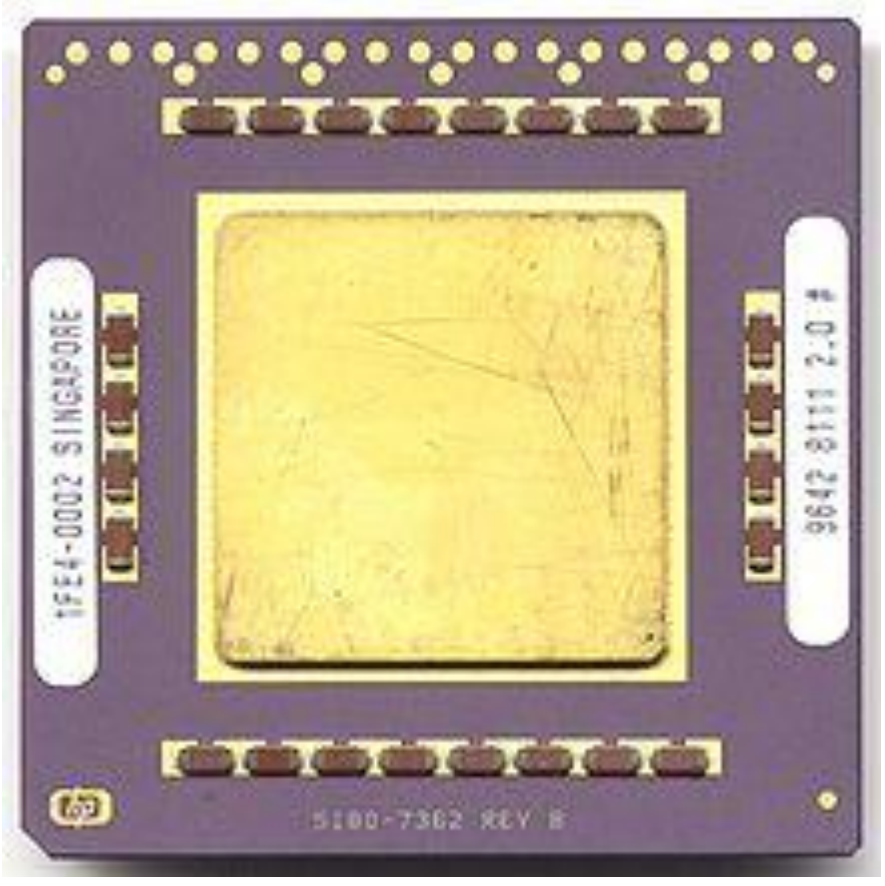

Figure 10 **HP 9000 PA-RISC Microprocessor**

26. **Performance Optimization with Enhanced RISC – Performance Computing (POWER-PC)[xiv]**. PowerPC is a RISC architecture created by the 1991 Apple–IBM–Motorola alliance, known as AIM.

a. Originally intended for personal computers design

b. Used in high performance processors.

c. PowerPC is largely based on IBM's earlier POWER architecture, and retains a high level of compatibility with it.

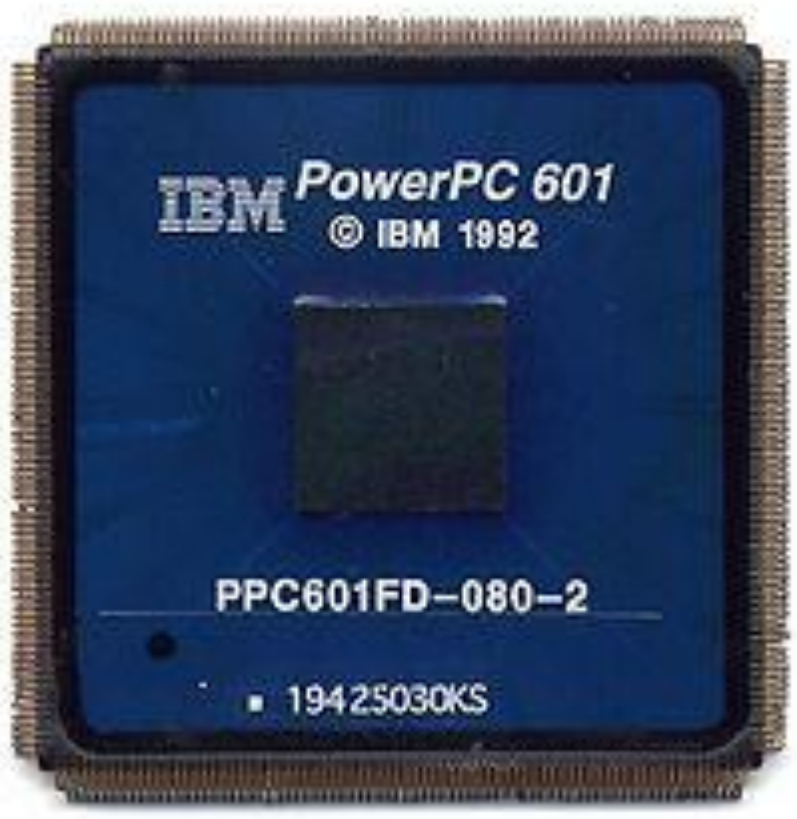

Figure 11 **PowerPC 600 Series developed by IBM Computer**

27. **SuperH[xv]**. SuperH (SH) is a 32-bit reduced instruction set computer (RISC) instruction set architecture (ISA) developed by Hitachi. It is implemented by microcontrollers and microprocessors for embedded systems. Its major categories are

a. SH-1. Used in microcontrollers for deeply embedded applications (CD-ROM drives, major appliances, etc)
b. SH-2. Used in microcontrollers with higher performance requirements, also used in automotive such as engine control units or in networking applications, and also in video game consoles, like the Sega Saturn. The SH-2 has also found home in a great many motor control applications.
c. SH-DSP. Initially developed for the mobile phone market, used later in many consumer applications requiring DSP performance for JPEG compression etc
d. SH-3. Used for mobile and handheld applications such as the Jornada, strong in Windows CE applications and market for many years in the car navigation market
e. SH-3 DSP. Used mainly in multimedia terminals and networking applications, also in printers and fax machines
f. SH-4. Used whenever high performance is required such as car multimedia terminals, video game consoles, or set-top boxes
g. SH-5. Used in high-end multimedia applications

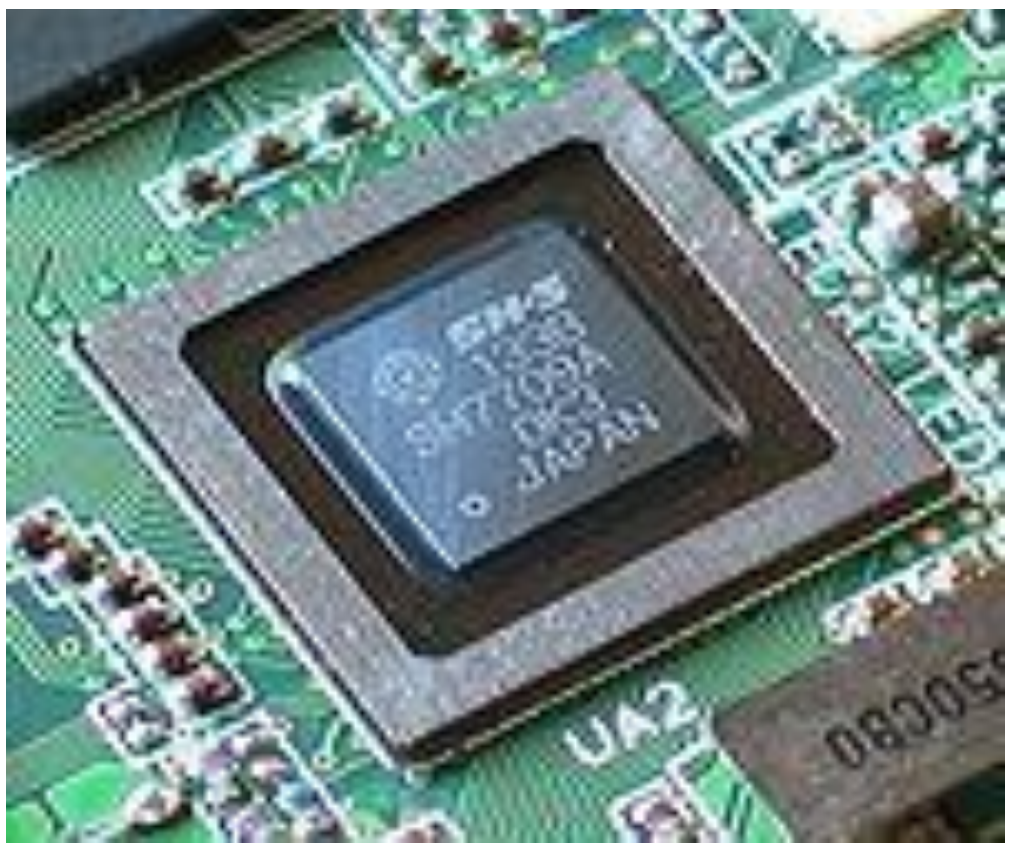

Figure 12 **SuperH Microprocessor**

28. **<u>Scalable Processor Architecture (SPARC)</u>**.[xvi] SPARC is a RISC instruction set architecture (ISA) developed by Sun Microsystems and introduced in 1986.

a. Implementations of the original 32-bit SPARC architecture were initially designed and used for Sun's Sun-4 workstation and server systems, replacing their earlier Sun-3 systems based on the Motorola 68000 family of processors.

b. Later, SPARC processors were used in servers produced by Sun Microsystems and designed for 64-bit operation.

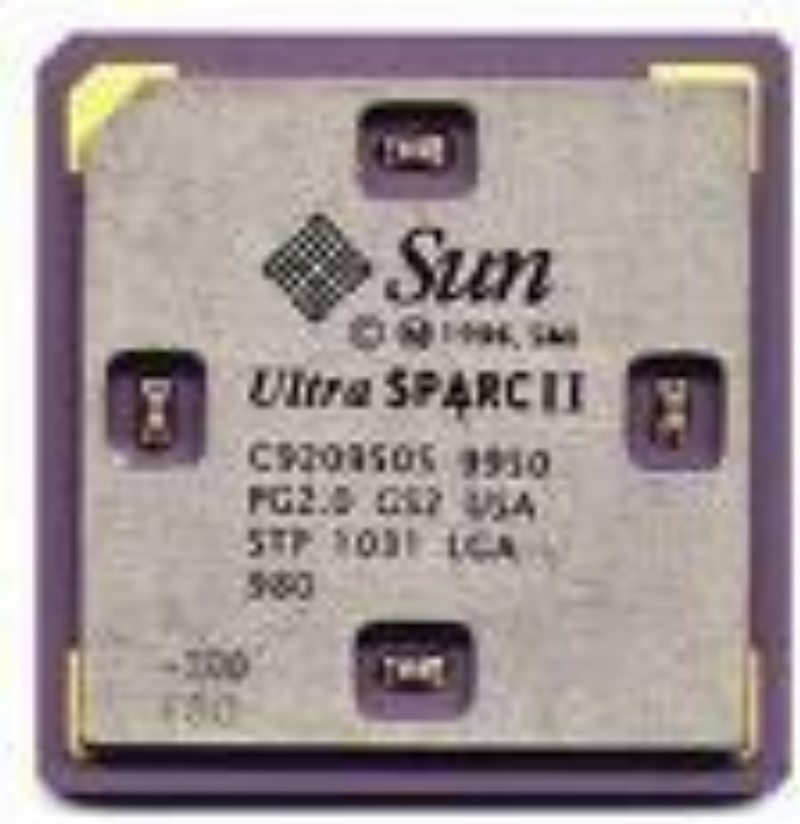

Figure 13 **<u>SPARC II Microprocessor developed by Sun</u>**

## <u>Conclusion</u>

29. We have introduced important characteristics that differentiate RISC designs from their CISC counterparts. CISC designs provide complex instructions and a large number of addressing modes. The rationale for this complexity is the desire to close the semantic gap that exists between high-level languages and machine languages. In the early days, effective usage of processor and memory resources was important. Complex instructions tend to minimize the memory requirements. Empirical data, however, suggested that compilers do not use these complex instructions; instead, they use simple instructions to synthesize complex instructions. Such observations led designers to take a fresh look at processor design philosophy. RISC principles, based on empirical studies on CISC processors, have been proposed as an alternative to CISC. Most of the current processor designs are based on these RISC principles.

______________________